\documentclass[aps,pra,twocolumn]{revtex4-1}

\usepackage{graphics}

\newcommand{\hata}[1]{\hat{a}_{#1}}
\newcommand{\hatb}[1]{\hat{b}_{#1}}
\newcommand{\hatx}[2]{\hat{X}_{#1}^{#2}}
\newcommand{\hatad}[1]{\hat{a}^{\dag}_{#1}}

\newcommand{\ket}[1]{| #1 \rangle}
\newcommand{\meanI}[1]{\langle #1 \rangle}
\newcommand{\var}[2]{\meanI{\Delta \hat{#1}_{#2}^{2}}}
\newcommand{\varI}[1]{\meanI{\Delta #1^{2}}}
\newcommand{\x}{\vec{x}}
\newcommand{\nquad}{\!\!\!\!\!\!}

\begin{document}
\title{Extracting Spatial Information from Noise Measurements
\\of Multi-Spatial-Mode Quantum States}
\author{A.~M. Marino}\email{alberto.marino@nist.gov}
\author{J.~B. Clark}
\author{Q. Glorieux}
\author{P.~D. Lett}
\affiliation{Quantum Measurement Division, National Institute of Standards and Technology, Gaithersburg, MD 20899 USA
\\and Joint Quantum Institute, NIST and the University of Maryland, Gaithersburg, MD 20899 USA}

\begin{abstract}
We show that it is possible to use the spatial quantum
correlations present in twin beams to extract information about the
shape of a mask in the path of one of the beams.  The scheme, based
on noise measurements through homodyne detection, is useful in the
regime where the number of photons is low enough that direct
detection with a photodiode is difficult but high enough that photon
counting is not an option.  We find that under some conditions the
use of quantum states of light leads to an enhancement of the
sensitivity in the estimation of the shape of the mask over what can be
achieved with a classical state with equivalent properties (mean
photon flux and noise properties).
In addition, we show that the level of enhancement that is obtained is a result of the quantum correlations and cannot be explained with only classical correlations.
\end{abstract}

\maketitle
\section{Introduction}

The field of quantum imaging studies the spatial distribution of the
quantum correlations in optical fields~\cite{Kolobov99,Kolobov07}.
It offers the possibility of improving the resolution of an imaging
system~\cite{Kolobov00}, enhancing image
detection~\cite{Brambilla08} or beam positioning~\cite{Treps03},
and implementing parallel quantum information
encoding~\cite{Bechmann00} and processing~\cite{Lassen07}.

An important goal of quantum imaging is to use the
multi-spatial-mode properties of quantum fields to extend the
enhancement in sensitivity that can be obtained with quantum states
from the temporal~\cite{Giovannetti11} to the spatial domain.  In this paper we
propose a scheme that achieves this goal. In particular, we show
that it is possible to take advantage of spatial quantum
correlations between two entangled beams to extract information
about the shape of a mask in the path of one of the beams.  We
consider a configuration based on noise measurements and homodyne
detection. As we are not extracting information from the mean field values, this technique becomes useful when the number of photons that is used to interrogate the mask is in the regime where it is low
enough that it cannot be readily detected with a photodiode, but
high enough that photon counting is not an option. Even though in
this regime the mask is interrogated by a small number of
photons, the use of homodyne detection makes it possible to perform
the measurement.

We analyze whether or not there is an advantage in the use of
quantum states  over classical ones with equivalent properties (mean
photon flux and single beam noise), and find that under some
conditions an enhancement of the sensitivity in the estimation of the shape
of the mask can be obtained with the quantum states.  We show
that the enhancement that is obtained is a result of the
quantum correlations and cannot be explained with only classical
correlations. Finally, we find that the level of enhancement increases with the number of spatial modes in the field probing the mask.

\section{Basic Configuration}

The possibility of extracting spatial information about a mask from
noise measurements of a beam that passes through it results from the
fact that in homodyne detection the shape of the local oscillator
(LO) effectively acts as a spatial filter that selects out the
portion of the field to be measured~\cite{Lugiato93,Lugiato97}.  That is, the spatial overlap
between the beam that goes through the mask and the LO affects the
efficiency of the measurement. Since the beam acquires the shape of
the mask after going through it, we can change the shape of the LO
when performing the homodyne detection to optimize  the measured
noise level and thus estimate the overlap.
Once the optimization criterion is met, the shape of the mask can be
deduced from the shape of the LO. In practice, as a result of the
type of measurement being done, this problem reduces to one of
parameter estimation. For this particular problem the quantity to
estimate is the overlap or mode matching between the LO and the beam that goes through the mask.

\begin{figure}[hpbt]
    \centering
    \includegraphics{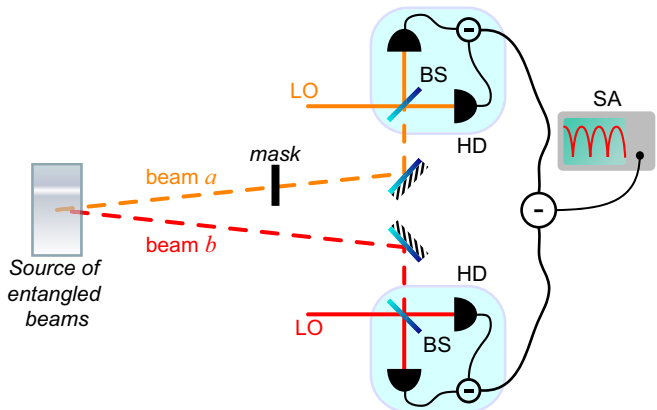}
    \caption{\footnotesize Configuration for estimating the shape of a mask through noise measurements of spatially multi-mode entangled beams. LO: local oscillator; BS: 50/50 beam splitter; HD: homodyne detector; SA: spectrum analyzer.}
\label{configuration}
\end{figure}

The basic configuration for such a scheme is shown in
Fig.~\ref{configuration}.  Entangled twin beams are used to probe
the mask whose shape we want to estimate. One of the beams goes
through the mask and is then detected with a homodyne detector while
the other one is directly sent to a second homodyne detector. The noise in the difference signal of the detectors is
used to estimate the overlap and thus extract information about the
shape of the mask.  When the entangled beams that are used contain
multiple spatial modes they also contain spatial quantum
correlations which are characterized by a coherence
area~\cite{Gatti04a,Brambilla04} that sets the spatial extent of the smallest
correlated areas between the two beams. If the beam contains
multiple coherence areas, and the size of these coherence areas is
smaller than the size of the mask, then the sensitivity in the
estimation can be enhanced.  Since a coherence area in one beam
is only correlated with a single corresponding coherence area in the
other beam~\cite{Boyer08a,Navez01}, it is important that the LOs of the two homodyne
detectors select out or measure the corresponding correlated regions
in the two beams.

In order to implement this scheme one would like to perform a series of noise measurements with a dynamically changing LO.  This could be done by using spatial light modulators to control the spatial profile of the LOs for beams $a$ and $b$. The shape of the LOs would be modified until the noise level is optimized, at which point the overlap between the LO of beam $a$ and the field that went through the mask is also optimized.  This implies that the shape of the LO of beam $a$ corresponds to the one of the field that went through the mask and makes it possible to infer the shape of the mask from the final shape of this LO.

This strategy is reminiscent of ghost imaging~\cite{Erkmen10}, in
which the goal is to image an object without it being in the path of
an imaging camera or detector. In ghost imaging the two beams that
are used are composed of streams of correlated photon pairs.  The
object is placed in the path of one of the beams, which is then
detected by a non-imaging, or ``bucket'', detector, which alone
reveals no information about the shape of the object.  The beam that
does not go through the mask is sent directly to a charge-coupled device (CCD) camera or
scanning point detector.  Correlations between the arrival of
photons on the detectors can then be used to reconstruct the object.
Even though originally it was believed that spatial quantum
correlations were required for ghost
imaging~\cite{Strekalov95,Pittman95,Abouraddy01}, it was later
realized that classical correlations could also be used for the same
purpose~\cite{Bennink02,Gatti04b}.  As we will show here, there are
two main differences between the scheme proposed here and ghost
imaging.  The first is that, with the proposed configuration,
information about the object can be obtained directly from
measurements of only the beam that goes through the mask.  The
second, and most important one, is that the use of quantum
correlations between the beams can lead to an enhancement of the
sensitivity in the estimation of the shape of the object over a classical
state with equivalent properties, and that this enhancement cannot
be explained assuming only classical correlations.

We compare the two-beam strategy presented in Fig.~\ref{configuration} with two analogous classically obtainable configurations.  In the first one we consider a single beam scheme in which the beam that goes through the mask shares the same photon flux and mode structure as the entangled beams in the two-beam strategy.  After going through the mask the beam is  sent to a homodyne detector. The resulting noise measurements are then used to estimate the shape of the mask.
Throughout the paper we refer to this scheme as the single-beam strategy.  In the second configuration we duplicate the two-beam strategy using classically correlated beams also sharing the same photon flux and mode structure as above.

\section{Multi-Spatial-Mode Properties \label{spatialmodes}}

In order to analyze both the two-beam and single-beam strategies
described above we need to take into account the multi-spatial-mode
nature of the fields, even for the case in which only one of the
spatial modes is not in a vacuum state. This can be done by
expanding the field operators in terms of a complete set of spatial
mode functions~\cite{Kolobov99}, such that
\begin{eqnarray}
    \hata{}(\x)&=&\sum_{i}A_{i}(\x)\hata{i}\label{modeexpansiona}\\
    \hatb{}(\x)&=&\sum_{i}A_{i}(\x)\hatb{i},
    \label{modeexpansionb}
\end{eqnarray}
where $A_{i}(\x)$ is an orthonormal and complete basis, $\x$ refers
to the transverse variables of the field, and $\hata{i}$ and
$\hatb{i}$ are the destruction operators for spatial mode $i$ for
beams $a$ and $b$, respectively.
We also assume that the state of the different spatial modes within each beam are not quantum correlated, such that the state of the field can be
written as~\cite{Martinelli03}
\begin{equation}
    \ket{\psi}=\ket{\psi_{1}}_{1}\cdots\ket{\psi_{N}}_{N}\ket{0}_{N+1}\cdots\ket{0}_{M},
    \label{state}
\end{equation}
where the first $N$ spatial modes of the field are excited with a wave function $\ket{\psi_{i}}$ for spatial mode $i$ and
the rest of the modes are in a vacuum state.  When more than one spatial mode is excited ($N>1$) the field is considered a multi-spatial-mode field.

\section{Shape Estimation through Noise Measurements}

The goal of the scheme described here is to estimate the field transfer
function of the mask, which is taken to be of the form
\begin{equation}
    t(\x)=\sqrt{T(\x)}e^{i\phi(\x)},
\end{equation}
where $T(\x)$ is the intensity transfer function and $\phi(\x)$ is the phase that a field would acquire after going through the mask and allows for the possibility of a phase mask. We can take into account the spatially dependent loss introduced by the mask by treating its effect on the field operator in the usual way.  That is, the field operators are transformed according to
\begin{eqnarray}
    \hata{out}(\x)&=&e^{i\phi(\x)}\left[\sqrt{T(\x)}\hata{in}(\x)+\sqrt{1-T(\x)}\hata{v}(\x)\right]\\
    \hatb{out}(\x)&=&\hatb{in}(\x),
\end{eqnarray}
where $\hata{v}(\x)$ is the corresponding operator for the vacuum field  that results from losses.  Note that in writing these expressions we have assumed that beam $a$ is the one that goes through the mask while beam $b$ has no losses.

As described above, we are interested in obtaining information about the transfer function of the mask through noise measurements of the signal that is obtained with balanced homodyne detection. As is typically done with this type of detection, we assume the LO to be a classical
field, such that the measured signal after beam $a$ goes through the mask is given by
\begin{eqnarray}
    \hat{I}_{a}&=&\int d\x\left[\hatad{out}(\x)\alpha_{LO}(\x)+\hata{out}(\x)\alpha_{LO}^{*}(\x)\right]
    \nonumber\\
    &=&\int d\x\left[e^{-i\phi(\x)}\sqrt{T(\x)}\alpha_{LO}(\x)\hatad{in}(\x)\right.\nonumber\\
    &&\left.+e^{-i\phi(\x)}\sqrt{1-T(\x)}\alpha_{LO}(\x)\hatad{v}(\x)+\rm{h.c.}\right],
    \label{homodyne}
\end{eqnarray}
where the integral is over the transverse profile of the fields, h.c. indicates the Hermitian conjugate, and $\alpha_{LO}(\x)$ is the spatial field profile of the LO. Similarly, for beam $b$ the measured signal is of the form given in Eq.~(\ref{homodyne}) with $T(\x)=1$ and $\phi(\x)=0$.

In order to simplify Eq.~(\ref{homodyne}) we use the mode expansion for the operators, Eq.~(\ref{modeexpansiona}), and expand the remaining product of functions in terms of the same basis, that is
\begin{eqnarray}
    e^{-i\phi(\x)}\sqrt{T(\x)}\alpha_{LO}(\x)&=&\alpha\sum_{i}t_{i}e^{i\phi_{i}}A_{i}(\x),\nonumber\\
    e^{-i\phi(\x)}\sqrt{1-T(\x)}\alpha_{LO}(\x)&=&\alpha\sum_{i}t_{i}'e^{i\phi_{i}'}A_{i}(\x),
    \label{funmodeexpansion}
\end{eqnarray}
where $\alpha$ gives the amplitude of the LO for beam $a$ and
$t_{i}e^{i\phi_{i}}$ and $t_{i}'e^{i\phi_{i}'}$ are the complex expansion coefficients for the product of functions in the selected basis.  In particular, $t_{i}$ and $\phi_{i}$ are the field overlap and relative phase between the LO, had it gone through the mask, and basis mode $i$.  Note that $t_{i}$ contains information about the shape of the LO and the amplitude transfer function of the mask, while $\phi_{i}$ contains information about the phase difference between the phase of the LO and any phase introduced by the mask. These expansions together with the orthonormality property of the basis allow us to simplify $\hat{I}_{a}$ to the form
\begin{equation}
    \hat{I}_{a}=\alpha\sum_{i}\left[t_{i}\hatx{a,i}{\phi}+t_{i}'\hatx{v,i}{\phi'}\right],
    \label{homodynesuma}
\end{equation}
where we have defined the generalized field quadrature for beam $a$ before the mask,
$\hatx{a}{\phi}=e^{i\phi}\hatad{in}+e^{-i\phi}\hata{in}$.  The subindex $v$ indicates that for those modes the field is in the vacuum state.

In the same way we can simplify the expression for the measured signal for beam $b$ by expanding the LO for beam $b$ in terms of the basis modes $A_{i}(\x)$, that is
\begin{equation}
    \alpha_{LO}(\x)=\alpha\sum_{i}\alpha_{i}e^{i\theta_{i}}A_{i}(\x),
    \label{LOmodeexpansion}
\end{equation}
where we have assumed the LO for beam $b$ has the same amplitude $\alpha$ as the one for beam $a$. In this expression $\alpha_{i}$ and $\theta_{i}$ are the field overlap and relative phase between the LO and basis mode $i$.  Note that we have defined the amplitude of the LO such that $\sum_{i}\alpha_{i}^{2}=1$.  We can now simplify $\hat{I}_{b}$ to the form
\begin{equation}
    \hat{I}_{b}=\alpha\sum_{i}\alpha_{i}\hatx{b,i}{\theta},
    \label{homodynesumb}
\end{equation}
where we have introduced the generalized field quadrature for the different spatial modes of beam $b$.

\subsection{Two-Beam Strategy}

We start by analyzing the two-beam strategy presented in
Fig.~\ref{configuration}. In this case entangled twin beams are
used to estimate the shape of the mask. For this strategy, the
signal to be measured is the noise in the difference signal, which
is given by
\begin{equation}
    M_{TB}=\varI{(\hat{I}_{a}-\hat{I}_{b})}_{N},
\end{equation}
where the $N$ subindex indicates that the variance or noise is normalized to the corresponding standard quantum limit (SQL) and the $TB$ subindex indicates the two-beam strategy.

Given that beam $a$ goes through the mask while beam $b$ sees no
losses, we find with the help of Eqs.~(\ref{homodynesuma}) and~(\ref{homodynesumb}) that
\begin{equation}
    \hat{I}_{a}-\hat{I}_{b}=\alpha\sum_{i}\left[t_{i}\hatx{a,i}{\phi}-\alpha_{i}\hatx{b,i}{\theta}+t_{i}'\hatx{v,i}{\phi'}\right],
\end{equation}
where we have assumed that the LOs for both beams $a$ and $b$ have the same amplitude $\alpha$ but not the same phase. Since we are only interested in the variance or noise of the signal, we can assume without loss of generality that the mean values for the quadratures of the field are zero, such that
\begin{equation}
    \meanI{\hat{I}_{a}-\hat{I}_{b}}=0.
\end{equation}
In addition, the situation when the mean value is zero is where the
scheme described here becomes more useful, as information about
the mask can only be obtained through noise measurements. With these
expressions and the assumption that the different spatial modes in each beam are not quantum correlated, we can show that the normalized noise in the difference signal is given by
\begin{eqnarray}
    M_{TB}&=&\frac{1}{2}+\frac{1}{2}\sum_{i}
    \left[t_{i}^{2}\varI{(\hatx{a,i}{\phi})}+\alpha_{i}^{2}\varI{(\hatx{b,i}{\theta})}\right.\nonumber\\
    &&\left.-2t_{i}\alpha_{i}\meanI{\hatx{a,i}{\phi}\hatx{b,i}{\theta}}-t_{i}^{2}\right],
    \label{mtb}
\end{eqnarray}
where we have used the relation $\sum_{i}t_{i}'^{2}=1-\sum_{i}t_{i}^{2}$ and have normalized the noise to the SQL for beams $a$ and $b$, that is, we divide by $2|\alpha|^{2}$.

\subsection{Single-Beam Strategy}

In order to consider the configuration in which only classical
resources are used, we analyze the single-beam strategy.  In this case
only the beam that goes through the mask is detected and the signal
to be measured is the normalized noise of that beam.  The measured signal for this strategy can be obtained from the one of the two-beam strategy, Eq.~(\ref{mtb}), by setting $\alpha_{i}=0$ and taking into account that for the single-beam strategy the normalization factor (the SQL) is given by $|\alpha|^{2}$ instead of $2|\alpha|^{2}$.  We thus have that
\begin{equation}
    M_{SB}=\var{I}{a}_{N}=\sum_{i}t_{i}^{2}\left[\varI{(\hatx{a,i}{\phi})}-1\right]+1,
    \label{Msb}
\end{equation}
where we have again assumed $\meanI{\hat{I}_{a}}=0$.

It is important to note a coherent state cannot be used with this strategy to extract information about the transfer function of
the mask, as the summation in Eq.~(\ref{Msb}) vanishes since a coherent state always has
$\varI{(\hatx{a,i}{\phi})}=1$.  As a result, $M_{SB}=1$ and no information can be obtained about the coefficients $t_{i}$ and thus about the shape of the mask. This is also the case for the two-beam strategy.  The reason that a coherent state is not useful is because  it has the same noise level as a vacuum state. As a result, the vacuum noise contribution that is introduced by the losses due to the mask will not change the noise level of the field after it goes through the mask.  Thus, this strategy requires a state whose noise level is initially different from that of a coherent state.  This is an important point since for this problem a coherent state does not provide a classical limit, as is usually the case with other estimation techniques.

\subsection{Shape Estimation}

As described above, the problem being considered reduces to one of
parameter estimation.  In particular, the parameter to estimate is the overlap between the LO and the beam that goes through the mask.  We can parameterize the noise of the signal that is measured with homodyne detection in terms of the transmission of the LO that one would obtain if it had gone through the mask instead, and make
it the parameter to estimate.  The transmission of the LO through
the mask is then given by
\begin{equation}
    T=\frac{\int d\x T(\x)|\alpha_{LO}(\x)|^{2}}{\int d\x|\alpha_{LO}(\x)|^{2}}
    =\sum_{i}t_{i}^{2},
    \label{Thomodyne}
\end{equation}
where the integral is over the transverse profile of the LO.  In obtaining this result we have used Eq.~(\ref{funmodeexpansion}) to express $T$ in terms of $t_{i}$.

For a general mask transfer function for which different transmission levels are possible, one would have to implement a different protocol in which a series of measurements is performed.  For example, one could use a series of LOs with different shapes, such as the basis modes,  in order to extract enough information to estimate the transfer function of the mask.  This would make it possible to extract all the  values of $t_{i}$.  Once all these values
are known and the input state is completely characterized it would be possible to reconstruct the complete transfer function of the mask.

For the case of a phase mask, information about the phase of the transfer function is contained in the phase factor of the expansion coefficients of Eq.~(\ref{funmodeexpansion}), that is $\phi_{i}$.  As can be seen from Eqs.~(\ref{mtb}) and~(\ref{Msb}) this phase factor appears in the generalized quadrature and selects the quadrature that is measured. For a phase sensitive state in which the noise properties depend on the phase being measured, one can modify the phase front of the LO for beam $a$ in such a way as to minimize or maximize the noise.  Once this is done the phase of the transfer function of the mask can be extracted from the phase front of the LO. For the two-beam strategy, which requires a LO for each beam, independent control of the phase front of each LO would be required, as the LO for beam $a$ will have to compensate for the additional phase that the beam acquires after going through the mask.
It should be noted that a phase insensitive state cannot be used to obtain information about a phase mask, as the measured noise will be independent of $\phi_{i}$ and thus will not provide information about the phase of the transfer function.

For the protocol as described here to work we require the intensity transfer function of the mask at each position $\x$ to have a transmission of either zero or one.  We assume that the area of the LO is the same as the one of the mask for $T=1$.  We also assume that the noise signal that is used for the estimation, $M_{TB}$ or $M_{SB}$, either grows or decreases as the transmission $T$ increases, such that a maximum or minimum is obtained when $T=1$.  Under these conditions once the noise level is optimized we know that the LO and the field that went through the mask have the same shape.  Thus, one can then obtain the intensity transfer function of the mask from the shape of the LO.

As can be seen from Eqs.~(\ref{mtb}) and~(\ref{Msb}) only spatial modes with a noise level different than the one of a coherent state contribute to the measurement.  As a result, even if only one spatial mode is not in a vacuum state, it is possible to obtain information about the transfer function of the mask. This means that a multi-spatial-mode field is not required for the estimation.  A single-mode field, however, is in general not detected with unit efficiency ($t_{i}^{2}\neq1$) when the LO has the same shape as the mask and $T=1$.  For an arbitrary mask, when the LO matches its shape, the expansion of the LO in terms of the basis modes requires more than a single mode. As we will show later, this reduction in efficiency will tend to lead to a reduction in the accuracy or sensitivity of the estimation of $T$ and having multiple spatial modes will make it possible to obtain a more accurate estimate of $T$.

\section{Sensitivity of the Estimation}

It is important to study the sensitivity that can be achieved in the estimation of the transmission $T$.  In general, the uncertainty in estimating parameter $P$ through
measurements of quantity $M$ is given by
\begin{equation}
    \Delta P^{2}=\frac{\Delta M^{2}}{|\partial M/\partial P|^{2}},
\end{equation}
which takes into account the slope of the measured signal with
respect to the parameter being estimated and the noise on the
measurement. As this uncertainty is reduced, the sensitivity of the
estimation is increased, thus making it possible to obtain a better
estimate for $P$.  We now calculate the sensitivity for both
strategies described above in the estimation of $T$.

\subsection{Two-Beam Strategy}

To determine the sensitivity of this strategy we need to calculate the noise of $M_{TB}$,
which means that we need the variance of the difference
noise signal, that is
\begin{equation}
    \Delta
    M_{TB}^{2}=\meanI{(\hat{I}_{a}-\hat{I}_{b})^{4}}-\meanI{(\hat{I}_{a}-\hat{I}_{b})^{2}}^{2}.
    \label{varnoisetb}
\end{equation}
To simplify this expression we assume that the fields that are used have Gaussian statistics, which means that higher order moments can be expressed in terms of first and second order moments~\cite{Loudon}.  In particular, this implies that for the quadrature of each mode
\begin{equation}
    \meanI{(\hatx{i}{\phi})^{4}}=3\meanI{(\hatx{i}{\phi})^{2}}^{2}.
    \label{gausstate}
\end{equation}
Using this property we find that Eq.~(\ref{varnoisetb}) can be
simplified to
\begin{eqnarray}
    \Delta M_{TB}^{2}&=&\frac{1}{2}\left\{1-T+\sum_{i}
    \left[t_{i}^{2}\varI{(\hatx{a,i}{\phi})}\right.\right.\nonumber\\
    &&\qquad\left.\left.+\alpha_{i}^{2}\varI{(\hatx{b,i}{\theta})}-2t_{i}\alpha_{i}\meanI{\hatx{a,i}{\phi}\hatx{b,i}{\theta}}\right]\vphantom{\sum_{i}}\right\}^{2}\nonumber \\
    &=& 2M_{TB}^{2}.
\end{eqnarray}

We thus find that the sensitivity in the estimation of $T$ with the two-beam strategy is given
by
\begin{eqnarray}
    \Delta T^{2}_{TB}&=&\frac{2M_{TB}^{2}}{\left|\partial M_{TB}/\partial T\right|^{2}}\nonumber\\
    &\nquad\nquad\nquad\nquad=&\nquad\nquad\frac{8M_{TB}^{2}}
    {\left|\frac{\partial\left[\sum_{i}
    (t_{i}^{2}\varI{(\hatx{a,i}{\phi})}+\alpha_{i}^{2}\varI{(\hatx{b,i}{\theta})}
    -2t_{i}\alpha_{i}\meanI{\hatx{a,i}{\phi}\hatx{b,i}{\theta}})\right]}{\partial
    T}-1\right|^{2}},\nonumber\\
    \label{Ttb}
\end{eqnarray}
where $M_{TB}$ is given by Eq.~(\ref{mtb}).  In general this expression needs to be evaluated numerically and its behavior depends on the mode structure of the entangled beams and how
the beam that goes through the mask is attenuated by it.

\subsection{Single-Beam Strategy}

To calculate the sensitivity of the single-beam strategy we need to calculate the variance of $M_{SB}$, which is given by
\begin{equation}
    \Delta M_{SB}^{2}=\meanI{\hat{I}_{a}^{4}}-\meanI{\hat{I}_{a}^{2}}^{2}.
\end{equation}
Using the mode expansion given in Eq.~(\ref{homodynesuma}) for
$\hat{I}_{a}$ and the Gaussian property of the state given in
Eq.~(\ref{gausstate}) we find that
\begin{eqnarray}
    \Delta M_{SB}^{2} &=&2\left[\sum_{i}t_{i}^{2}\varI{(\hatx{a,i}{\phi})}+1-T\right]^{2}\nonumber\\
   &=&2M_{SB}^{2}.
\end{eqnarray}
We thus have that the sensitivity for the single-beam strategy is given by
\begin{equation}
    \Delta T^{2}_{SB}=\frac{2M_{SB}^{2}}{\left|\partial M_{SB}/\partial T\right|^{2}}
    =2\frac{\left[\sum_{i}t_{i}^{2}\varI{(\hatx{a,i}{\phi})}+1-T\right]^{2}}
    {\left|\frac{\partial\left(\sum_{i}t_{i}^{2}\varI{(\hatx{a,i}{\phi})}\right)}{\partial T}-1\right|^{2}}.
    \label{Tsb}
\end{equation}
We again find that an analytical solution as a function of $T$ is not possible in general.  Note that this result is also valid for quantum states such as single-mode squeezed states.  Even though using this type of state might lead to an enhanced sensitivity when compared to the same strategy using classical states, we will only consider classical states with the single-mode strategy in the rest of the paper.

\subsection{Enhancement in Sensitivity with Quantum States}

The uncertainty in the estimation of $T$ given in Eqs.~(\ref{Ttb})
and~(\ref{Tsb}) are general results that apply to any state with the
only restriction being that it have Gaussian statistics.  In general
whether or not an enhancement in sensitivity can be obtained with
the two-beam strategy using quantum states over the single-beam
strategy using classical states depends on the actual mode structure
and shape of the mask. The functional form of these results will change for each configuration and, in general, a numerical analysis would be required to compare the two strategies.

In order to compare the two strategies we consider the limit in which all the spatial modes have the same noise properties, that is, $\varI{(\hatx{a,i}{\phi})}=\varI{(\hatx{a}{\phi})}$, $\varI{(\hatx{b,i}{\theta})}=\varI{(\hatx{b}{\theta})}$, and   $\meanI{\hatx{a,i}{\phi}\hatx{b,i}{\theta}}=\meanI{\hatx{a}{\phi}\hatx{b}{\theta}}$.
This situation would correspond to the limit in which the state is composed of an infinite number of spatial modes all with the same properties. Even though this is not a realistic limit, it is still instructive to consider it, as it allows us to obtain
an analytical solution for Eqs.~(\ref{Ttb}) and~(\ref{Tsb}) and illustrate the enhancement that can be obtained through the use of quantum states.

For the two-beam strategy we assume that the quantum state of
light used is given by vacuum twin beams.  For this type of state
each beam by itself is phase insensitive, so we can drop the phase
dependence ($\varI{(\hatx{a}{\phi})}=\var{X}{a}$ and $\varI{(\hatx{b}{\theta})}=\var{X}{b}$). In addition, the
noise properties of both individual beams are the same, that is
$\var{X}{b}=\var{X}{a}$.  In this case Eq.~(\ref{mtb}) reduces to
\begin{eqnarray}
    M_{TB}&=&1+\frac{1}{2}\left[\varI{(\hatx{a}{})}-1\right](T+1)\nonumber\\
    &&-\left[\varI{(\hatx{a}{})}-M_{0}^{\phi,\theta}\right]\sum_{i}t_{i}\alpha_{i},
\end{eqnarray}
where we have written $\meanI{\hatx{a}{\phi}\hatx{b}{\theta}}$ in terms of the noise properties of the input beams (before the mask) and defined the input normalized noise of the quadrature difference  $M_{0}^{\phi,\theta}=\varI{(\hatx{a}{\phi}-\hatx{b}{\theta})}_{N}$.
For the correct choice of $\phi$ and $\theta$, $M_{0}^{\phi,\theta}$ corresponds to the amount of quadrature difference squeezing in the twin beams~\cite{Boyer08b}.  In order to simplify this result further we assume that the shape of the LO for beam $b$ is not changed and that it selects out only one spatial mode, that is, $\alpha_{1}=1$ and $\alpha_{i\neq1}=0$. Note that this will not lead to optimum results, as the LO for beam $b$ will not be matched to the one for beam $a$.  This will lead to the detection of  uncorrelated coherence areas between beams $a$ and $b$. Finally, we assume that for the spatial mode of beam $a$ corresponding to the one that was selected by the LO in beam $b$ the field transmission through the mask scales linearly with the transmission.  As an example we take $t_{1}=0.8T$.

For the single-beam strategy we consider a classical state with
the same single beam noise and mean photon
flux as the beam that goes through the mask in the two-beam
configuration.  Such a state corresponds to a thermal state and is
exactly the state that is obtained from the vacuum twin beams when
only one of the beams is measured~\cite{Loudon}.  Even though this state is
obtained from an entangled state, it can be produced with a
classical source and can be considered a classical state.  For such
a state under the conditions described above we find that
\begin{equation}
    \Delta T^{2}_{SB}=2\frac{(T\var{X}{a}+1-T)^{2}}
    {(\var{X}{a}-1)^{2}}.
    \label{Dtsbsimp}
\end{equation}

\begin{figure}[hpbt]
    \centering
    \includegraphics{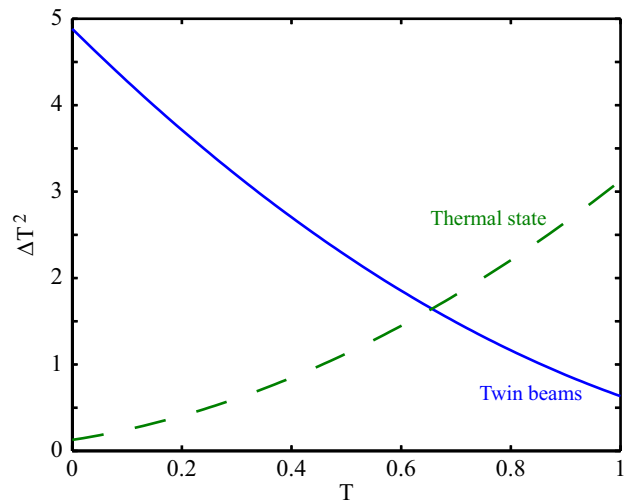}
    \caption{Uncertainty in the estimation of the transmission ($\Delta T^{2}$) as a function of the transmission $T$ for a thermal state (single-beam strategy) and twin beams (two-beam strategy) for $\var{X}{a}=5$ and $M_{0}=0.1$.}
\label{homdynesens}
\end{figure}

Figure~\ref{homdynesens} compares the uncertainty in the estimation of $T$ for both strategies in the case in which the phases of the LOs are set to measure the minimum noise in the quadrature difference with a value of $M_{0}=0.1$.  For a minimum uncertainty state this implies a normalized single beam noise $\var{X}{a}=5$.
As we can see from this figure, the two-beam strategy with twin beams offers an enhancement in sensitivity over the single-beam one with a thermal state for larger values of $T$ ($T>0.65$ for the particular example shown in Fig.~\ref{homdynesens}).  The value of $T$ at which such an enhancement occurs depends on the initial noise properties of the states and the functional form assumed for $t_{1}$. The largest enhancement is obtained for $T=1$ which is where we want to operate, since it is at this value of $T$ where the LO and the field that went through the mask have the same shape and an estimation of the shape of the mask can be made.

We now consider the more favorable case for the two-beam strategy in which the LOs for beams $a$ and $b$ only measure correlated coherence areas.  If we assume that the mode structure for beams $a$ and $b$ is the same and that quantum correlations are only present between modes with the same spatial profile, then this case implies that the LO for beam $b$ has the same shape as the LO for beam $a$ after the mask. This will only be the case for values of $T\approx1$.  For this
region of operation, having the same shape for both LOs implies that $\alpha_{i}\approx t_{i}$, which allows us to simplify Eq.~(\ref{Ttb}) to the form
\begin{equation}
    \Delta T^{2}_{TB} \approx 2 \frac{\left[(1-T)(\var{X}{a}+1)+2TM_{0}\right]^{2}}
    {\left[2M_{0}-(\var{X}{a}+1)\right]^{2}}.
\end{equation}
In this case, the sensitivity for the single-beam strategy is still given by Eq.~(\ref{Dtsbsimp}). In the limit of large levels of squeezing ($\var{X}{a}\gg1\gg M_{0}$) and $T=1$ we find that the sensitivity is enhanced by a factor
\begin{equation}
    \frac{\Delta T_{SB}^{2}}{\Delta T_{TB}^{2}}\rightarrow \left(\frac{\var{X}{a}}{M_{0}}\right)^{2},
\end{equation}
which shows that the sensitivity enhancement factor grows with the level of entanglement or squeezing present in the twin beams.

\subsection{Enhancement due to the Multi-Spatial-Mode Nature of the Field}

As we have shown, a multi-spatial-mode field is not required to estimate the transfer function of the mask, as a field with only a single excited spatial mode can provide enough information for the estimation.  Next,  we study whether there is any advantage to the use of a multi-spatial-mode field by reducing the problem to a one dimensional one and calculating the uncertainty in the estimation of $T$ numerically. To do so we consider a square aperture with a transmission of one inside the square and zero everywhere else.  We assume that the LOs for both beams have the same shape and size as the mask with a constant amplitude, such that the goal is to find the position of the mask.  The problem is reduced to a one dimensional one by limiting the movement of the LO in only one direction.

To perform the numerical analysis we assume a Hermite-Gauss basis for the mode decomposition and calculate the enhancement in sensitivity obtained with the two-beam strategy over the single-beam strategy, $\Delta T_{SB}^{2}/\Delta T_{TB}^{2}$, as a function of the number of excited spatial modes $N$ in the field.  We assume that all the modes that are excited have the same noise properties as before, $\var{X}{a}=\var{X}{b}=5$ and $M_{0}=0.1$.
\begin{figure}[hpbt]
    \centering
    \includegraphics{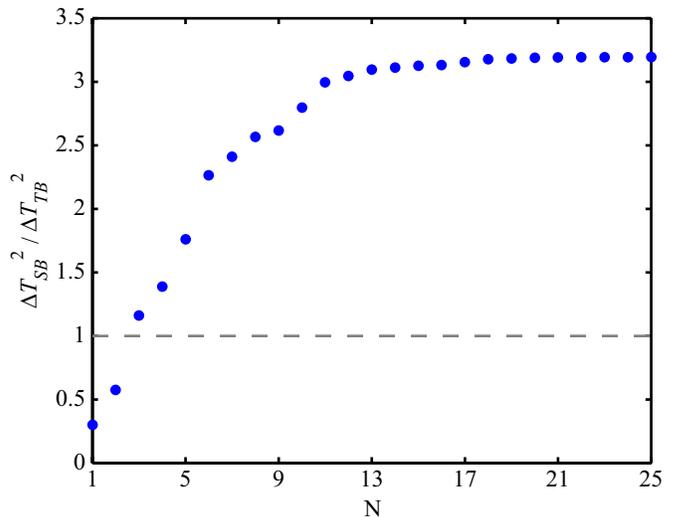}
    \caption{\footnotesize Enhancement in sensitivity of the estimation of $T$ of the two-beam strategy with respect to the single-beam as a function of the number of excited spatial modes $N$.  An enhancement is obtained for $\Delta T_{SB}^{2}/\Delta T_{TB}^{2}>1$.}
\label{simulation}
\end{figure}
Figure~\ref{simulation} shows the enhancement in sensitivity as a function of $N$ for $T=1$. This value of the transmission corresponds to the one at which one would operate to estimate the shape of the mask.  In this figure a value of $\Delta T_{SB}^{2}/\Delta T_{TB}^{2}>1$ indicates an enhancement in the estimation of $T$ with the two-beam strategy based on a quantum state over the one-beam strategy based on a classical state. Due to the symmetry of the basis modes and the aperture, the odd Hermite-Gauss modes do not contribute to the estimation of $T$.  As a result, only even terms are taken into account in the numerical analysis. In obtaining this result we have added the modes from lower to higher order.  That is, $N=1$ contains only mode TEM$_{00}$, $N=2$ adds mode TEM$_{02}$, $N=3$ adds mode TEM$_{20}$, and so on until all the even modes between TEM$_{00}$ and TEM$_{88}$ have been added for $N=25$. For the particular mode structure and mask we have used we find that for one and two spatial modes the classical strategy actually does better than the quantum one.  For a larger number of modes, however, we do get an enhancement with quantum states.  Additionally, we find that as the number of modes increases the level of enhancement in sensitivity also increases and saturates at an enhancement factor of around 3.2.  This result shows that as the number of spatial modes present in the field that probes the mask increases so does the enhancement that can be obtained with quantum states over classical ones.

\subsection{Role of Quantum Correlations}

A natural question to ask is whether the enhancement that can be
obtained with entangled states is due to the quantum correlations or
if it can also be obtained through classical correlations.  To
answer this question we need to consider the two-beam strategy
in the presence of only classical correlations.  That is, we assume
that the state of the field is described by the density matrix~\cite{Duan00a}
\begin{equation}
    \hat{\rho}=\sum_{i}p_{i}\hat{\rho}_{a,i}\otimes\hat{\rho}_{b,i},
    \label{rhoclass}
\end{equation}
where $\hat{\rho}_{a,i}$ and $\hat{\rho}_{b,i}$ are the density matrices for beams $a$ and $b$, respectively, and $p_{i}$ allows for a statistical mixture of separable states.
This expression makes it possible to consider classical correlations.

For the state of the field given by Eq.~(\ref{rhoclass}) we have
that
\begin{equation}
    \meanI{\hatx{a,i}{\phi}\hatx{b,i}{\theta}}
    =\sum_{j}p_{j}\meanI{\hatx{a,i}{\phi}}_{j}\meanI{\hatx{b,i}{\theta}}_{j}=0.
\end{equation}
With this result, the sensitivity in estimating $T$ given by
Eq.~(\ref{Ttb}) reduces to
\begin{equation}
    \Delta T^{2}_{NQ}=2\frac{\left[\sum_{i}
    \left(t_{i}^{2}\varI{(\hatx{a,i}{\phi})}+\alpha_{i}\varI{\hatx{b,i}{\theta})}\right)+1-T\right]^{2}}
    {\left|\frac{\partial\left(\sum_{i}
    t_{i}^{2}\varI{(\hatx{a,i}{\phi})}\right)}{\partial
    T}-1\right|^{2}},
    \label{Tnc}
\end{equation}
where the subindex $NQ$ indicates that there are no quantum correlations between beams $a$ and $b$ and we have assumed $\alpha_{i}$ to be independent of $T$.
Since $\varI{(\hatx{b,i}{\theta})}\geq0$ we can see by comparing
Eq.~(\ref{Tnc}) with Eq.~(\ref{Tsb}) that $\Delta
T^{2}_{NQ}\geq\Delta T^{2}_{SB}$.  That is, the presence of only
classical correlations in the two-beam strategy does not make it
possible to obtain an enhancement in sensitivity over the classical
single-beam strategy and in fact, in general, it leads to a decrease in
sensitivity. This result shows that quantum correlations are needed
in order to obtain the enhancement in sensitivity.

\section{Conclusion}

We have shown that it is possible to use noise measurements to
estimate the transfer function of a mask in the path of an optical beam. For this purpose we have analyzed two different strategies: a two-beam configuration based on quantum states and a one-beam configuration based on classical states.  We have shown that an enhancement can be obtained with the use of quantum states over classical ones.  This enhancement is a result of the quantum correlations present between the two beams and cannot be explained through the presence of only classical correlations.

Even though a multi-spatial-mode field is not required for the estimation of the transfer function in either the two-beam or single-beam configurations, we have shown that the presence of multiple spatial modes can increase the enhancement in sensitivity that is obtained with the quantum states.  In addition, the level of enhancement increases with the level of entanglement or squeezing between the twin beams used for the estimation.

Although we have limited the analysis to the case of a mask with a transmission value at a given point of either zero or one, we have shown that it is possible to extend the analysis presented to obtain information about a phase mask and to treat the case of soft apertures.  This makes it possible to estimate
the complete transfer function of an arbitrary mask through noise measurements.

This work was supported by the US Air Force Office of Scientific Research.

\end{document}